\documentclass[amsmath,amssymb,twocolumn]{revtex4}
\usepackage[parfill]{parskip}    
\usepackage{graphicx}
\usepackage{amssymb}
\usepackage{epstopdf}
\usepackage{color}
\usepackage{hyperref}
\usepackage{graphicx}
\usepackage{pdfpages}

\begin{document}
\title{\bf Five-dimensional scale-dependent black holes with constant curvature and Solv horizons}
\author{E. Contreras$^1$
{
\footnote{econtreras@yachaytech.edu.ec}
} 
, 
\'A. Rinc\'on$^2$ 
{
\footnote{arrincon@uc.cl}
}
and P. Bargue\~no$^3$
{
\footnote{p.bargueno@uniandes.edu.co}
}
}
\affiliation{$^1$ School of Physical Sciences \& Nanotechnology, Yachay Tech University, 100119 Urcuqu\'i, Ecuador.
\\
$^2$ Instituto de F\'{\i}sica, Pontificia Universidad Cat\'olica de Chile,  
Av. Vicu\~na Mackenna 4860, Santiago, Chile.
\\
$^3$ Departamento de F\'{\i}sica,
Universidad de los Andes, Apartado A\'ereo {\it 4976}, Bogot\'a, Distrito Capital, Colombia.}
\begin{abstract}
In this work, we investigate five-dimensional scale-dependent black hole solutions by modelling their event horizon
with some of the eight Thurston three-dimensional geometries. Specifically, we construct constant curvature scale-dependent 
black holes and also the more exotic scale-dependent Solv black hole. These new solutions are obtained by promoting both 
the gravitational and the cosmological couplings to $r$-dependent functions, in light of a particular description of 
the effective action inspired by the high energy philosophy.
Interestingly, the so-called running parameter, together with the topology of the
event horizon, control the asymptotic structure of the solutions found. Finally, differences in both the entropy and the 
temperature between the classical and the scale-dependent Solv black hole are briefly commented.
\end{abstract}
\maketitle
\section{Introduction}

Topological techniques in General Relativity \cite{Penrosebook, Hawkingbook,Gerochbook}, 
developed mainly by Penrose, Hawking and Geroch, reveal their power in the celebrated singularity theorems (see, for example,
\cite{Hawkingbook}). Among the most important ingredients of these theorems are the energy conditions and topological 
considerations on certain three dimensional hypersurfaces. Interestingly, the interplay between topology and energy
conditions also appear in Hawking's first black hole topology theorem \cite{Hawking1972}, which asserts that the event horizon
of an asymptotically flat stationary four-dimensional black hole obeying the dominant energy condition is a two-sphere.
Of the various ways to escape this theorem, we can mention going to higher dimensions \cite{Horowitzbook} 
or considering matter sources that violate the dominant energy condition.

When considering five-dimensional spacetimes, the event horizon can be, in principle, any compact orientable three-dimensional
manifold. However, due to the Thurston geometrization conjecture proved by Perelman \cite{Perelman2002}, 
the event horizon can be endowed with a metric locally isometric to one of the eight Thurston geometries \cite{Thurstonbook}.
Within these geometries, the simples one are the Euclidean space $\mathbb{E}^3$, the three-sphere $\mathbb{S}^3$, the
hyperbolic space $\mathbb{H}^3$ and the products $\mathbb{S}^1\times \mathbb{S}^2$ and $\mathbb{S}^1\times \mathbb{H}^2$.
There are also three non-trivial geometries called the Solv geometry, the Nil geometry and the geometry of the universal cover of
SL$_2(\mathbb{R})$.

In Ref. \cite{Cadeau2001}, the authors catalogued solutions to the five-dimensional vacuum Einstein equations which were
modelled on three-dimensional geometries of spherical, hyperbolic, flat or product type. In addition, they found two families
of new black hole solutions modelled by the Solv and the Nil geometries. Some of these results were generalized in 
Ref. \cite{Hassaine2015}, where new Nil black holes with hyperscaling violation were studied. 
Even more, in the context of AdS/CMT, thermoelectric transport coefficients form charged Solv and Nil black holes have
been recently reported \cite{Arias2017}. In addition, DC conductivities have been computed for Solv, Nil and 
SL$_2(\mathbb{R})$ black branes \cite{Arias2018}. Finally, we note that Solv and Nil solutions electromagnetically charged through a 
dilatonic source have been considered in Ref. \cite{Bravo2018}.

Interestingly, all these five-dimensional black hole solutions have been found, to the best of our knowledge, 
only within classical General
Relativity. Therefore, it is our interest to model, if possible, the event horizon of certain black holes beyond General Relativity
by some of the eight Thurston three-geometries. As a preliminary step in order to attack more general cases, in this
work we will consider black hole solutions embedded in the so-called scale-dependent gravity \cite{Contreras:2013hua,Koch:2013rwa,Rodrigues:2015hba,Koch:2015nva,
Koch:2016uso,Rincon:2017goj,Rincon:2017ypd,Rincon:2018sgd,
Contreras:2018dhs,Rincon:2018dsq,Contreras:2018gct,Contreras:2018gpl,
Rincon:2018lyd,Contreras:2018swc,Rincon:2019cix,Contreras:2017eza,Rincon:2019zxk} 
which, as it is well known, has become an alternative tool to introduce semiclassical corrections in black hole 
solutions in $2+1$ and $3+1$ dimensional space-times.

The main idea in which the scale-dependent scenario is inspired can be summarized as follows: following the lessons of 
Weinberg's Asymptotic Safety program, we consider that the effective action ({\it i. e.}, a modified version of the classical 
action, which includes quantum-mechanical corrections) is the fundamental object and, therefore, the corresponding equations
of motion should be derived using this action. Irrespectively of how complicated these effective actions could be, a common feature 
between them is always present: the couplings involved acquire a scale-dependence. Taking this statement seriously, 
we shall analyse how the aforementioned black holes suffer deviations with respect to their classical counterparts.

The manuscript is organized as follows: Section \ref{scale_setting} summarizes the main points of the scale-dependent
gravitational theory. In Section \ref{classical}, some classical black holes in five dimensions with constant curvature and
Solv horizons are reviewed in order to extend them, in Section \ref{BlackHoleSolution}, into the scale-dependent theory.
Section \ref{thermo} shows the scale-dependent modifications in both the temperature and entropy of the scale-dependent
Solv black hole with respect to its classical counterpart, and concluding remarks and some comments are given in 
Section \ref{final}.

\section{Scale-dependent gravity \label{scale_setting}}

In order to be able to include any quantum correction in certain black hole solutions, it is mandatory to specify the object 
which describes the ``fundamental theory". Analogously to standard gravity (where the classical action is taken to be the 
main object), scale-dependent gravity takes advantage of the idea provided by the Asymptotic Safety program in which the 
gravitational effective action, $\Gamma[k,g_{\mu \nu}, \cdots]$, where $k$ is a scale field, describes the theory. 
One of the most remarkable features in quantum field theories is that the effective action for the gravitational field, 
indeed at low-energy, acquires a scale dependence. This effect appears at the level of the couplings which means that it runs 
according to certain energy scale, this fact being a generic result of quantum field theory.

%
%
%

The Asymptotic Safety program, which is based on a non-trivial ultra-violet fixed point for the leading dimensionless 
gravitational couplings is, by far, where these ideas have been best implemented. Let us point out that it
was Weinberg, in his seminal work \cite{Weinberg:1976xy}, who introduced this program. 
Substantial improvement has been made up to now \cite{Wetterich:1992yh,Morris:1993qb,Reuter:1996cp,Reuter:2001ag,Litim:2002xm,Litim:2003vp,Niedermaier:2006wt,Niedermaier:2006ns,Gies:2006wv,Machado:2007ea,Percacci:2007sz,Codello:2008vh,Benedetti:2009rx,Manrique:2009uh,Manrique:2010am,Manrique:2010mq,Eichhorn:2010tb,Litim:2011cp,Falls:2013bv,Dona:2013qba,Falls:2014tra,Eichhorn:2018yfc,Eichhorn:2017egq}.

%
During the last years, scale-dependent gravity has
been used to construct black hole backgrounds both by improving classical solutions with the scale dependent couplings from 
Asymptotic Safety \cite{Bonanno:1998ye,Bonanno:2000ep,Emoto:2005te,Bonanno:2006eu,Reuter:2006rg,Koch:2007yt,Hewett:2007st,Litim:2007iu,Burschil:2009va,Falls:2010he,Casadio:2010fw,Reuter:2010xb,Cai:2010zh,Falls:2012nd,Becker:2012js,Koch:2013owa,Koch:2014cqa,Gonzalez:2015upa,Torres:2017ygl,Pawlowski:2018swz}
and by solving the  gap equations of a generic scale-dependent action 
\cite{Contreras:2013hua,Koch:2013rwa,Rodrigues:2015hba,Koch:2015nva,
Koch:2016uso,Rincon:2017goj,Rincon:2017ypd,Rincon:2018sgd,Contreras:2018dhs,
Rincon:2018dsq,Contreras:2018gct,Contreras:2018gpl,Rincon:2018lyd,
Contreras:2018swc,Rincon:2019cix,Contreras:2017eza,Rincon:2019zxk}.
This last approach has revealed certain non-trivial features regarding the black hole entropy and the energy conditions.
Even more, more recently, scale-dependent regular black holes \cite{Contreras:2017eza} and traversable (vacuum) 
scale-dependent wormholes \cite{Contreras:2018swc} have been found, showing that, in some sense and in particular situations,
scale-dependent gravity might shed light on how to cure, in an effective way, some of the classical problems related to
singularities and the appearance of exotic matter inside wormholes. From the cosmological point of view,
the impact of scale dependence has been studied in Refs. 
\cite{Bonanno:2001xi,Weinberg:2009wa,Tye:2010an,Bonanno:2010bt,Bonanno:2001hi,Koch:2010nn,Grande:2011xf,Copeland:2013vva,Bonanno:2015fga,Bonanno:2017gji,Hernandez-Arboleda:2018qdo,Bonanno:2018gck,Canales:2018tbn}.

The simplest formulation of scale-dependent gravity is given by the 
scale-dependent version of the Einstein-Hilbert action as follows
%
%
\begin{eqnarray}\label{action}
\Gamma[g_{\mu\nu},k]=\int \mathrm{d}^{n}x\sqrt{-g}
\bigg[\frac{1}{2 \kappa_{k}} \bigg(R-2\Lambda_{k}\bigg)\bigg],
\end{eqnarray}
where $k$ is a scale-dependent field related to a renormalization scale,   
$\kappa_k \equiv 8 \pi G_k$ is the Einstein coupling,
and $G_{k}$ and $\Lambda_{k}$ refer to
the scale-dependent gravitational and cosmological couplings, respectively.
The modified Einstein's equations are obtained by taking variations with respect to the metric field $g_{\mu\nu}$:
\begin{eqnarray}\label{einstein}
G_{\mu\nu}+g_{\mu\nu}\Lambda_{k}=-\Delta t_{\mu\nu},
\end{eqnarray}
where the so-called non-matter energy-momentum tensor, $\Delta t_{\mu\nu}$, is defined according to
\cite{Reuter:2003ca,Koch:2010nn}
\begin{eqnarray}\label{nme}
\Delta t_{\mu\nu}=G_{k}\Bigl(g_{\mu\nu}\square -\nabla_{\mu}\nabla_{\nu}\Bigl)G_{k}^{-1}.
\end{eqnarray}
%
By taking the variation of the effective action with respect to the scale field, $k(x)$, one
imposes~\cite{Koch:2014joa}
\begin{align}\label{scale}
\frac{\mathrm{d}}{\mathrm{d} k} \Gamma[g_{\mu \nu}, k] =0,
\end{align}
which can be seen as an a posteriori condition towards background independence
\cite{Stevenson:1981vj,Reuter:2003ca,Becker:2014qya,Dietz:2015owa,Labus:2016lkh,Morris:2016spn,Ohta:2017dsq}.
The $\beta$-functions describe the renormalization group running of both Newton's constant and the cosmological coupling. 
These $\beta$-functions are, in general, unknown because they depend of how we solve the problem. Given that we do not want 
to compute them,
we do not have enough information in order to find both the metric and the scale field. In order to bypass this issue, here we consider
that the couplings $G_k$ and $\Lambda_k$ inherit the dependence on the space-time coordinates from the space-time dependence of
the scale field, $k(x)$. Therefore, these couplings are written as $G(x)$ and $\Lambda(x)$
\cite{Koch:2014joa,Rincon:2017goj}. This idea, together with an appropriate choice for the line element, allows, in principle, to solve the problem in situations with a high degree of symmetry.

\section{Some classical five-dimensional black hole solutions}\label{classical}

In this section we will briefly review some of the five-dimensional black holes obtained within Einstein's theory in 
Ref. \cite{Cadeau2001}. This choice of solutions is related to those which we have been able to generalize into the scale-
dependent theory, as subsequent sections will show.

The classical Einstein-Hilbert action is, in five dimensions, given by
\begin{align}\label{classical_action}
I_0[g_{\mu\nu}] &= \int \mathrm{d}^{5}x \sqrt{-g}
\bigg[
\frac{1}{2 \kappa_0} 
\bigg(R - 2\Lambda_0 \bigg)
\bigg],
\end{align}
where $\kappa_0 \equiv 8 \pi G_0$ is the gravitational coupling, $G_0$ is
Newton's constant and $\Lambda_0$ is the cosmological coupling. 
Varying the classical action \eqref{classical_action} yields the equations of motion, i.e., 
\begin{align}
R_{\mu \nu} - \frac{1}{2}R g_{\mu \nu} &= -\Lambda_0 g_{\mu \nu},
\end{align}
which can be expressed as
\begin{equation}
R_{ab}=\frac{2 \Lambda_{0}}{3}g_{ab}.
\end{equation}
If the metric is written as
\begin{equation}
ds^2 = -V(r)dt^2 + V(r)^{-1}dr^2 + f^2(r)dx^{i}dx^{j},
\end{equation}
where $d\tilde s^2= dx^{i}dx^{j}$ is the metric induced on the $t=$ constant, $r=$ constant surfaces, 
two type of solutions are obtained, 
depending on the integration of $f''(r)$ (see \cite{Cadeau2001} for details).
The first set of the solutions are given by \cite{Cadeau2001}
\begin{eqnarray}
\label{constant}
&&f(r)=r \nonumber \\
&&V(r)=k-\frac{2G_{0}M}{r^2}-\frac{r^2\Lambda_{0}}{6},
\end{eqnarray}
and
\begin{eqnarray}
&&d\tilde s^2 = d\xi^2+ \sin^{2}\xi d\Omega^2 \; (k =1) \nonumber \\
&&d\tilde s^2 = d\xi^2+ \sinh^{2}\xi d\Omega^2 \; (k = - 1) \nonumber \\
&&d\tilde s^2 = d\xi^2+ d\theta^2 + d \phi^2 \; (k = 0) .
\end{eqnarray}
These solutions are a warped product between a certain two-dimensional spacetime and a three-dimensional 
Riemannian manifold of constant curvature (spherical, hyperbolic or flat). 
Interestingly, there is also a second set of solutions which are 
products of a two-dimensional spacetime with an Einstein three-manifold, with no warping. The interested reader can consult
the specific form of these solutions in \cite{Cadeau2001}.
%
%
%
In adition to these constant-curvature black hole solutions, Solv and Nil geometries were found in Ref. \cite{Cadeau2001}. 
In particular, the Solv black hole is found for $\Lambda_{0}<0$ and it has a metric given by
%

\begin{align}
\label{solv}
\mathrm{d}s^2 = - B(r) \mathrm{d}t^2 + B(r)^{-1} \mathrm{d}r^2 + \frac{3}{(-\Lambda_0)}\mathrm{d}\tilde{s}^2,
\end{align}
assuming the following definitions:
\begin{align}
B(r) \equiv & -\frac{2}{9}\Lambda_{0} r^2 - \frac{2\sqrt{G_{0}M}}{r} 
\\
d\tilde s^2 = & r^2\left(e^{2 z}dx^2+e^{-2 z}dy^2\right)+dz^2.
\end{align}

%
As previously commented, we are interested in the generalization of some classical black hole solutions. Specifically, 
the solutions given by Eqs. (\ref{constant}) and (\ref{solv}) will be generalized in the next section
when certain effective deformation of the classical theory is incorporated in terms of scale-dependent gravity.

\section{Five-dimensional scale-dependent AdS black holes\label{BlackHoleSolution}}

Due to the symmetries we want to explore, let us make the choice $G=G(r)$ and $\Lambda=\Lambda(r)$ for the quasi-dynamical Newton and cosmological couplings, respectively. It is remarkable that, in non-trivial situations, for instance with the Kerr black hole, 
the functional dependence of the coupling could be more complex. In our particular case, 
we arrive to the following vacuum equations:
\begin{equation}
G_{\mu\nu} + g_{\mu\nu}\Lambda(r) + \Delta t_{\mu \nu}(r)
=0,    
\end{equation}
In the next subsections we shall consider several horizon geometries.
\\
\\
\subsection{Black holes with spherical and hyperbolic horizons}

Here we consider line elements parametrized as
\begin{eqnarray}
    ds^2&=&-V(r)dt^2+\frac{dr^2}{V(r)}+ f(r)^2\left(d\xi^2+ \sin^2 \xi d\Omega^2\right) \nonumber \\
    ds^2&=&-V(r)dt^2+\frac{dr^2}{V(r)}+ f(r)^2\left(d\xi^2+ \sinh^2 \xi d\Omega^2\right) \nonumber ,\\
\end{eqnarray}
where $d\Omega^2$ is the round metric for the two-sphere.

We get the following solutions:
\\
\begin{widetext}
\begin{eqnarray}
&&f(r)=r \nonumber \\
&&V(r)=k-\frac{2 G_{0} M}{r^2}-\frac{r^2 \Lambda_{0}}{6}+
\frac{8 G_{0} M \epsilon }{3 r}-k\frac{2 r \epsilon }{3}-4 G_{0} M \epsilon ^2 \nonumber \\
&&+k\frac{5 r^2 \epsilon ^2}{9}+8 G_{0} M r \epsilon ^3 -\left(-\frac{2}{3}k r^2 \epsilon ^2+8 G_{0} M r^2 \epsilon ^4\right)\log(1+\frac{1}{\epsilon r}) \nonumber \\
&&G(r)=\frac{G_{0}}{1+\epsilon r} \nonumber \\
&&\Lambda(r)=\frac{1}{9 r^2 (1+r \epsilon )^2}\left(-12 G_{0} M \epsilon ^2 (-1+2 r \epsilon  (4+3 r \epsilon  (11+10 r \epsilon )))
\right) \nonumber\\
&&+\frac{1}{9 r^2 (1+r \epsilon )^2}r^2 \left(\epsilon ^2 k (27-10 r \epsilon  (2+5 r \epsilon ))+3 (1+r \epsilon ) (3+5 r \epsilon ) \Lambda_{0}\right)\nonumber \\
&&+\frac{1}{9 r^2 (1+r \epsilon )^2} \left(
12 r^2 \epsilon ^2 (1+r \epsilon ) (3+5 r \epsilon ) \left(-k+12 G_{0} M \epsilon ^2\right)\right)\nonumber \\
&&\frac{1}{9 r^2 (1+r \epsilon )^2} \left(\log(1+\frac{1}{\epsilon r})\right),
\end{eqnarray}
\end{widetext}
where $k=\pm 1$ stands for the curvature of the horizon.

It is worth mentioning that 
the scale-dependent system can be integrated without providing some extra information, in contrast to which occurs in 
lower dimensional cases \cite{Contreras:2013hua,Koch:2013rwa,Rodrigues:2015hba,
Koch:2015nva,Koch:2016uso,Rincon:2017goj,Rincon:2017ypd,
Rincon:2018sgd,Contreras:2018dhs,Rincon:2018dsq,Contreras:2018gct,
Contreras:2018gpl,Rincon:2018lyd,Contreras:2018swc,Rincon:2019cix,Contreras:2017eza,Rincon:2019zxk}. 
To be more precice, in $2+1$ and $3+1$-dimensional space-times, the scale-dependent system is undeterminated: 
we have more unknowns that equations to be solved, so decreasing the degrees of freedom is mandatory. In all of these cases, 
the null energy condition plays an important role in the sense that it allows to decrease these number of degrees of freedom.
In particular, this condition leads to a differential equation for the Newton coupling, $G(r)$, which is finally 
written as $G(r) = G_{0}/(1 + \epsilon r)$.

The fact that the Newton coupling obtained for the five-dimensional solutions here presented coincides with that of 
lower-dimensional cases is not a coincidence. It was proved that, after a suitable choice of certain null vector, 
higher-dimensional solutions are forced to obtain the same gravitational coupling \cite{Rincon}. This is true indeed without 
invoking the null energy condition or another suitable constraint. Therefore, 
the information encoded into the null energy condition is contained into the modified Einstein field equations.

Finally, we note that in the limit $\epsilon\rightarrow0$ the classical (non-running) solution is recovered. Specifically,
\begin{eqnarray}
\label{ksolclass}
\lim_{\epsilon\rightarrow 0}f(r) &=& r \nonumber \\
\lim_{\epsilon\rightarrow 0}V(r) &=& k-\frac{2 G_{0} M}{r^2}-\frac{r^2 \Lambda_{0}}{6} \nonumber \\
\lim_{\epsilon\rightarrow 0} G(r) &=& G_{0} \nonumber \\
\lim_{\epsilon\rightarrow 0}\Lambda(r) &=& \Lambda_{0}.
\end{eqnarray}

\subsubsection{Asymptotics}

In order to both interpretate the results and evaluate the differences in some thermodynamics quantities between
the classical and the running solutions, here we show the asymptotic expansion, for small $\epsilon$ and large $r$,
for both the Ricci curvature and the lapse function. They read
\begin{eqnarray}
&&R = \frac{10 \Lambda_0}{3}+\left(-\frac{16 G_0 M}{3 r^3}+k \frac{8}{r}\right) \epsilon \nonumber \\
&&+\frac{2}{9} 
\left(-k 23+\frac{108 G_0 M}{r^2}+k 60 \log (r \epsilon)\right) \epsilon ^2+\mathcal{O}[\epsilon ]^3 \nonumber \\
&&R= \frac{10}{9} \left(-k 10 \epsilon ^2+3 \Lambda_0\right)+k\frac{2}{r^2}+\mathcal{O}\left[\frac{1}{r}\right]^3,
\end{eqnarray}
and
\begin{eqnarray}
&&V(r)= \left(k-\frac{2 G_0 M}{r^2}-\frac{r^2 \Lambda_0}{6}\right)+\left(\frac{8 G_0 M}{3 r}-k\frac{2 r}{3}\right) \epsilon 
\nonumber \\
&&+\frac{1}{9} \left(-36 G_0 M+k 5 r^2-k 6 r^2 \log [\epsilon r]\right) \epsilon ^2+\mathcal{O}[\epsilon ]^3\nonumber\\
&&V(r)=\left(\frac{k 5 \epsilon ^2}{9}-\frac{\Lambda_0}{6}\right) r^2+ \frac{2 k}{3}+\frac{2k}{9 \epsilon  r}
-\frac{k}{6 \epsilon ^2 r^2}\nonumber \\
&&-\frac{2 \left(-k+12 G_0 M \epsilon ^2\right)}{15 \epsilon ^3 r^3}+\frac{-k+12 G_0 
M \epsilon ^2}{9 \epsilon ^4 r^4} \nonumber \\
&&+\mathcal{O}\left[\frac{1}{r}\right]^5.
\end{eqnarray}
From these results, a couple of comments are in order. (i) Both the spherical ($k=1)$ and hyperbolic ($k=-1$) solutions give
place to an effective cosmological constant given by
\begin{equation}
\Lambda_{\mathrm{eff}}=\frac{5 k}{9}\epsilon^2-\frac{\Lambda_{0}}{6}.
\end{equation}
Interestingly, only in the spherical case, this $\Lambda_{\mathrm{eff}}$ can be turned off for certain values of the running
parameter. (ii) A mass-like term appears only in the spherical case. Specifically,
\begin{equation}
G_{0}M_{\mathrm{eff}}=\frac{1}{12\epsilon^2}.
\end{equation}
(iii) A charge-like term appears in both cases. Specifically
\begin{equation}
G_{0}Q^2_{\mathrm{eff}}=\frac{12G_{0}M\epsilon^2-k}{9\epsilon^4}.
\end{equation}
As for the effective cosmological constant, this $Q_{\mathrm{eff}}$ can be turned off for certain values of the running
parameter. And (iv), both the spherical and hyperbolic solutions incorporate a defect-like term (monopole-like
 \cite{Barriola1989}) with an effective
deficit angle $\delta_{\mathrm{eff}}$ given by the term
\begin{equation}
\delta_{\mathrm{eff}}\sim \frac{2}{3}k.
\end{equation}
We note that this deficit angle does not depend on the running parameter. It is worth mentioning that this is
not a surprising finding in the context of scale-dependent gravity. In fact, in Ref. \cite{Koch:2015nva} the authors 
reported an $\epsilon$-independent deficit angle in the asymptotic behaviour of the line element at infinity in 
the scale-dependent Einstein-Maxwell system.

Even more, it is noticeable that all these effective quantities which emerge at infinity do not turn off when the running
parameter goes to zero. As commented in \cite{Koch:2015nva}, we ascribe this behaviour to 
the fact all dimensionless terms of the form $\epsilon r$ are incompatible with first taking the limit $r \rightarrow \infty$
and then $\epsilon \rightarrow 0$. Clearly, if the limit of small $\epsilon$ is taken first, the classical results are
recovered, as shown in Eq. (\ref{ksolclass}).

\subsection{Black holes with planar horizon}

For a line element parametrized as
\begin{align}
    ds^2=-V(r)dt^2+\frac{dr^2}{V(r)}+ f(r)^2\left(d\xi^2+ d\theta^2 + d\phi^2 \right),
\end{align}
we get the following solution:
\\
\begin{widetext}
\begin{eqnarray}
&&f(r)=r \nonumber \\
&&V(r)=-\frac{2 G_{0} M}{r^2}-\frac{r^2 \Lambda_{0}}{6}+
\frac{8 G_{0} M \epsilon }{3 r}-4 G_{0} M \epsilon ^2 \nonumber \\
&&-\frac{5 r^2 \epsilon ^2}{9}+8 G_{0} M r \epsilon ^3 -8 G_{0} M r^2 \epsilon ^4 \log(1+\frac{1}{\epsilon r}) \nonumber \\
&&G(r)=\frac{G_{0}}{1+\epsilon r} \nonumber \\
&&\Lambda(r)=
\frac{1}{3 r^2 (1+r \epsilon )^2}\left(-4 G_0 M \epsilon ^2 (-1+2 r \epsilon  (4+3 r \epsilon  (11+10 r \epsilon )))\right) \nonumber \\
&&+\frac{1}{3 r^2 (1+r \epsilon )^2}\left(r^2 (1+r \epsilon ) (3+5 r \epsilon ) \Lambda_0\right) \nonumber \\
&&+\frac{1}{3 r^2 (1+r \epsilon )^2}\left(
+48 G_0 M r^2 \epsilon ^4 (1+r \epsilon ) (3+5 r \epsilon ) 
\log \left[1+\frac{1}{\epsilon r }\right]\right), \nonumber
\end{eqnarray}
\end{widetext}
Also in this case the classical (non-running) solution is recovered when the running parameter is turned off. That is,
\begin{eqnarray}
\lim_{\epsilon\rightarrow 0}f(r) &=& r \nonumber \\
\lim_{\epsilon\rightarrow 0}V(r) &=& -\frac{2 G_{0} M}{r^2}-\frac{r^2 \Lambda_{0}}{6} \nonumber \\
\lim_{\epsilon\rightarrow 0}G(r) &=& G_{0} \nonumber \\
\lim_{\epsilon\rightarrow 0}\Lambda(r) &=& \Lambda_{0}.
\end{eqnarray}
\subsubsection{Asymptotics}

In this planar case, the expansions of both $R$ and $V(r)$ are given by
\begin{eqnarray}
R &=& \frac{10 \Lambda_0}{3}-\frac{16 (G_0 M) \epsilon }{3 r^3}+\frac{24 G_0 M \epsilon ^2}{r^2}+\mathcal{O}[\epsilon ]^3\nonumber \\
R&=& \frac{10 \Lambda_0}{3}-\frac{8 (G_0 M)}{3 \epsilon ^2 r^6}+\mathcal{O}\left[\frac{1}{r}\right]^7,
\end{eqnarray}
and
\begin{eqnarray}
&&V(r)= \left(-\frac{2 G_0 M}{r^2}-\frac{r^2 \Lambda_0}{6}\right)+\frac{8 G_0 M \epsilon }{3 r} \nonumber \\
&&-4 (G_0 M) \epsilon ^2+\mathcal{O}[\epsilon ]^3\nonumber \\
&&V(r)= -\frac{\Lambda_0 r^2}{6}-\frac{8 (G_0 M)}{5 \epsilon  r^3}+\frac{4 G_0 M}{3 \epsilon ^2 r^4}
+\mathcal{O}\left[\frac{1}{r}\right]^5,
\end{eqnarray}
showing that the rule of the running parameter can be interpreted in this case as related to 
an effective charge, which is given by
\begin{equation}
Q^2_{\mathrm{eff}}=\frac{4 M}{3 \epsilon^2}.
\end{equation}

\subsection{Solv black holes}

When appropriate units are used, a five-dimensional metric with three-dimensional Solv geometry can be parametrized as
\cite{Cadeau2001}
{\small
\begin{equation}
    ds^2=-V(r)dt^2+\frac{dr^2}{V(r)}+ A e^{2 z} r^2 dx^2 + A e^{-2 z} r^2 dy^2 + A dz^2.
\end{equation}
}
Using this ansatz, we get the following solution:
{\small
\begin{eqnarray}
\label{solvscale}
&&A=-\frac{3}{\Lambda_{0}} \nonumber \\
&&V(r)=-\frac{2 \sqrt{G_0 M}}{r}+3 \sqrt{G_0 M} \epsilon -6 \sqrt{G_0 M} r \epsilon ^2-\frac{2 r^2 \Lambda_0}{9} \nonumber \\
&&+6 \sqrt{G_0 M} r^2 \epsilon ^3 \log \left[1+\frac{1}{r \epsilon }\right]
\nonumber \\
&&G(r)=\frac{G_{0}}{1+\epsilon r} \nonumber \\
&&\Lambda(r)=\frac{1}{3 A r (1+r \epsilon )^2}\left(9 A \sqrt{G_0 M} \epsilon ^2+108 A \sqrt{G_0 M} r \epsilon ^3 \right) \nonumber \\
&&+\frac{1}{3 A r (1+r \epsilon )^2}\left(108 A \sqrt{G_0 M} r^2 \epsilon ^4 \right) \nonumber\\
&&+ \frac{1}{3 A r (1+r \epsilon )^2}\left(r (1+r \epsilon ) (-3+2 A \Lambda_0+r \epsilon  (-3+4 A \Lambda_0))\right)
\nonumber \\
&&-\frac{1}{3 A r (1+r \epsilon )^2} \left(54 A \sqrt{G_0 M} 
r \epsilon ^3 (1+r \epsilon ) (1+2 r \epsilon ) \nonumber \right)\\
&& -\frac{1}{3 A r (1+r \epsilon )^2} \left(\log \left[1+\frac{1}{r\epsilon }\right]\right).
\end{eqnarray}
}
We note that in the limit $\epsilon\rightarrow0$ the classical (non-running) solution is recovered:
\begin{eqnarray}
\lim_{\epsilon\rightarrow 0}A&=& -\frac{3}{\Lambda_{0}} \nonumber \\
\lim_{\epsilon\rightarrow 0}V(r)&=&-\frac{2 \sqrt{G_{0} M}}{r}-\frac{2 r^2 \Lambda_{0}}{9}\equiv V_{0}(r)\nonumber \\
\lim_{\epsilon\rightarrow 0}G(r)&=&G_{0} \nonumber \\
\lim_{\epsilon\rightarrow 0}\Lambda(r)&=&\Lambda_{0}.
\end{eqnarray}
\subsubsection{Asymptotics}

In the Solv case, the corresponding expansions are written as

\begin{eqnarray}
R &=& \frac{10 \Lambda_0}{3}-\frac{6 \sqrt{G_0} M \epsilon }{r^2}+\frac{36 \sqrt{G_0 M} \epsilon ^2}{r}
+\mathcal{O}[\epsilon ]^3  \nonumber\\ 
R &=& \frac{10\Lambda_0}{3}-\frac{12 \sqrt{G_0 M}}{5 \epsilon ^2 r^5}+\mathcal{O}\left[\frac{1}{r}\right]^6.
\end{eqnarray}

\begin{eqnarray}
\label{asym}
V(r)&=& -\frac{2 \left(9 \sqrt{G_0 M}+r^3 \Lambda_0\right)}{9 r}+3 \sqrt{G_0 M} \epsilon \nonumber \\
&-&6 \left(\sqrt{G_0 M} r\right) 
\epsilon ^2+\mathcal{O}[\epsilon ]^3\nonumber \\
V(r)&=&-\frac{2 \Lambda_0 r^2}{9}-\frac{3 \sqrt{G_0 M}}{2 \epsilon  r^2}
+\frac{6 \sqrt{G_0 M}}{5 \epsilon ^2 r^3}\nonumber \\
&&-\frac{\sqrt{G_0 M}}{\epsilon ^3 r^4}+\mathcal{O}\left[\frac{1}{r}\right]^5.
\end{eqnarray}
In this case, the running parameter gives place asymptotically to an effective charge given by
\begin{equation}
Q_{\mathrm{eff}}=\frac{3}{2}\frac{\sqrt{G_{0}M}}{\epsilon}.
\end{equation}
Interestingly, due to the sign of the term which goes with $r^{-4}$, 
this interpetation is only valid when $\epsilon<0$ (see eq. (2.3) of Ref. \cite{Arias2017} for charged Solv black holes).

Therefore, the combination of the running parameter, $\epsilon$, together with
the considered topology ($\mathbb{E}^3$, $\mathbb{S}^3$, $\mathbb{H}^3$ and Solv), give place to different asymptotic structures
for the solutions here reported. Specifically, as shown in this section and summarized in Table \ref{table1},
an effective cosmological constant (different to $\Lambda_{0}$) is obtained for non-planar constant curvature horizons. 
On the contrary, both planar and Solv-geometries retain the classical cosmological
constant. Even more, although effective Maxwellian charges appear in all the solutions here reported, only the scale-dependent
Solv black hole contains a Reissner-Nordstr\"om-like structure without gravitational monopoles, which are, on the contrary,
included, in both the $\mathbb{S}^3$ and $\mathbb{H}^3$ cases. Finally, we have noted that the planar case is the only type which
does not give place to a mass-like term. It is important to remark that the asymptotic structure here outlined has not to be
taken literally. In fact, the asymptotic limits of the solutions here presented include other terms which are difficult
to interpretate and which typically fall-off slower than those we have given an interpretation in terms of an
effective mass, charge, monopole or cosmological constant.

\begin{table}[h!]
  \begin{center}
    \caption{Asymptotics of classical and improved (scale-dependent) black holes. See text for details.}
    \label{table1}
    \begin{tabular}{l|c|r} 
      \textbf{Geometry} & \textbf{Classical} & \textbf{Improved}\\
      \hline
      $\mathbb{E}^3$ & $(M,\Lambda_{0})$ & $(Q_{\mathrm{eff}},\Lambda_{\mathrm{eff}})$ \\
      $\mathbb{S}^3$ & $(M,\Lambda_{0})$ & $(\delta_{\mathrm{eff}},M_{\mathrm{eff}},Q_{\mathrm{eff}},\Lambda_{\mathrm{eff}})$ \\
      $\mathbb{H}^3$ & $(M,\Lambda_{0})$ & $(\delta_{\mathrm{eff}},M_{\mathrm{eff}},Q_{\mathrm{eff}},\Lambda_{\mathrm{eff}})$ \\
      Solv           & $(M,\Lambda_{0})$ & $(M_{\mathrm{eff}},Q_{\mathrm{eff}},\Lambda_{\mathrm{eff}})$\\
    \end{tabular}
  \end{center}
\end{table}

\section{Entropy and temperature for the scale-dependent Solv black hole}\label{thermo}

In order to evaluate the differences in some thermodynamical quantities introduced by the scale-dependence procedure, only
the more exotic scale-dependent Solv black holes will be considered. The extension of these results to scale-dependent
black holes with constant curvature horizons it is straightforward and it will no be treated in this work.

Let us define the volume element of the compact Solv spacetime as
\begin{equation}
\Omega_{\mathrm{Solv}}=\int_{\Omega_{x}\times\Omega_{y}\times\Omega_{x}}\sqrt{A} dx\, dy\, dz,
\end{equation}
where the $\Omega_{I}$'s for $I=(x,y,z)$ stand for the compact ranges of the horizon coordinates, $x_{I}$.

On one hand, as shown in Refs. \cite{Arias2017,Bravo2018}, the Bekenstein-Hawking entropy for classical (non-running) Solv black 
holes is given by
\begin{equation}
\label{entropy}
S_{0}=\frac{2\pi}{\kappa} |\Omega_{\mathrm{solv}}|\sqrt{\frac{2}{3}}r_{h}^2,
\end{equation}
where we remind the reader that $\kappa=8 \pi G_{0}$ and $r_{h}$ stands for the event horizon corresponding to Eq. (\ref{solv}).

On the other hand, although the scale-dependent Solv black hole has a different event horizon compared to its non-running
counterpart, we arrive to an improved entropy very similar to that given by Eq. (\ref{entropy}) given by
\begin{equation}
\label{entropya}
S=\frac{2\pi}{8\pi G(r_{H})} |\Omega_{\mathrm{solv}}|\sqrt{\frac{2}{3}}r_{H}^2,
\end{equation}
where $r_{H}$ stands for the event horizon corresponding to Eq. (\ref{solvscale}).

A simple expression which permits to compare the entropies in both the classical and the running situations is
\begin{equation}
\label{entropybis}
S= S_{0}(1+ \epsilon r_{H})\left(\frac{r_{H}}{r_{h}}\right)^2.
\end{equation}
Concerning the temperature, after employing the well-known formula
\begin{equation}
T=\frac{1}{4\pi}\left|\lim_{r\rightarrow r_{+}}\frac{\partial_{r} g_{tt}}{\sqrt{-g_{tt}g_{rr}}}\right|,
\end{equation}
where $r_{+}$ refers to the event horizon for the classical or the running solution, we get that
\begin{equation}
T=T_{0}\frac{V'(r_{H})}{V_{0}'(r_{h})}.
\end{equation}
As commented in previous works \cite{Contreras:2013hua,Koch:2013rwa,Rodrigues:2015hba,Koch:2015nva,
Koch:2016uso,Rincon:2017goj,Rincon:2017ypd,Rincon:2018sgd,Contreras:2018dhs,
Rincon:2018dsq,Contreras:2018gct,Contreras:2018gpl,Rincon:2018lyd,
Contreras:2018swc,Rincon:2019cix,Contreras:2017eza,Rincon:2019zxk}, 
the running parameter has to be smaller than the other scales entering the problem. 
Therefore, taking the asymptotic expansion
for $V(r)$ obtained in Eq. (\ref{asym}) up to second order in $\epsilon$, we get that the new event horizon lies at
\begin{equation}
r_{H}=r_{h}-\frac{1}{2}\epsilon r_{h}^2 -\epsilon^2 r_{h}^3+ \mathcal{O}[\epsilon]^3.
\end{equation}
Within this approximation, the corrections to both the temperature and the entropy are found to be
\begin{equation}
T = T_{0}\left[1-2 \epsilon r_{h}-\frac{7}{4}\epsilon^2 r_{h}^2 \right]+ \mathcal{O}[\epsilon]^3,
\end{equation}
and 
\begin{equation}
\label{scorr}
S=S_{0}\left[1-\frac{13}{4}r_{h}^2\epsilon^2\right]+ \mathcal{O}[\epsilon]^3.
\end{equation}
Finally, it is interesting to note that the corrected temperature includes a term linear in $\epsilon$ which is absent in
the corresponding correction for the entropy. In this sense, the entropy remains more robust with respect to
the scale-dependence corrections than the temperature, which responds faster to the effects of the running.


\section{Concluding remarks}\label{final}

In this work we have constructed, for the first time, five dimensional scale-dependent black holes. Although we have been
able to generalize some of the classical (non-running) solutions presented in Ref. \cite{Cadeau2001}, it remains
to be seen if all the classical black holes with Thurston geometries have their scale-dependent counterparts. 
Interestingly, all the solutions found incorporate the same running for the Newton coupling without invoking the null 
energy condition, in contrast with some previous works \cite{Contreras:2013hua,Koch:2013rwa,Rodrigues:2015hba,Koch:2015nva,
Koch:2016uso,Rincon:2017goj,Rincon:2017ypd,Rincon:2018sgd,Contreras:2018dhs,
Rincon:2018dsq,Contreras:2018gct,Contreras:2018gpl,Rincon:2018lyd,
Contreras:2018swc,Rincon:2019cix,Contreras:2017eza,Rincon:2019zxk}.
It is remarkable that, at lower dimensions, the null energy condition supplements the set of equations to be solved. 
However, 
at higher dimensions, this requirement is not mandatory. Despite of that, given the symmetry of the problem, the Schwarzschild 
ansatz is consistent with the null energy condition as was previously reported in \cite{Koch:2016uso,Rincon}.
Interestingly, after comparing the result here obtained for the running gravitational coupling with the
corresponding results provided by the Asymptotic Safety program \cite{Wetterich:1992yh,Morris:1993qb,Reuter:1996cp,Reuter:2001ag,Litim:2002xm,Litim:2003vp,Niedermaier:2006wt,Niedermaier:2006ns,Gies:2006wv,Machado:2007ea,Percacci:2007sz,Codello:2008vh,Benedetti:2009rx,Manrique:2009uh,Manrique:2010am,Manrique:2010mq,Eichhorn:2010tb,Litim:2011cp,Falls:2013bv,Dona:2013qba,Falls:2014tra,Eichhorn:2018yfc,Eichhorn:2017egq} one finds that a
matching is straight forward for  the scale setting choice  $k(z)\sim z$. As noted in \cite{Rincon:2019cix}, this choice seems peculiar, 
since in a naive scale setting one usually expects $k \sim 1/z$ for dimensional reasons. However, it is important to note that
this finding is not in contradiction with the Asymptotic Safety 
paradigm itself but with the naive scale setting $k\sim 1/z$, which is only an educated guess applied a 
posteriori, when one tries
to improve classical solutions with the running couplings found in Asymptotic Safety. 
After briefly discussing the effects due to the combination of both the topology
of the horizon and the scale-dependence, we have computed the corrections to the temperature and the entropy for the
scale-dependent Solv black hole, finding linear and quadratic growing behaviours with the running parameter and recovering the
classical results when the theory reduces to General Relativity. To conclude, we would like to mention that this work is, to 
the best of our knowledge, the first attempt to construct black holes with Thurston horizons in theories beyond General Relativity.

\section*{Acknowledgments}
The author P. B. was supported by the Faculty of Science and Vicerrector\'{\i}a de Investigaciones of 
Universidad de los Andes, Bogot\'a, Colombia. The author A. R. was supported by the CONICYT-PCHA/Doctorado Nacional/2015-21151658.
P. B. dedicates this work to Ana Bargue\~no-Dorta.

\end{document}